\documentclass[12pt]{article}

\usepackage[margin=1in,a4paper,nohead,dvips]{geometry}
\usepackage{amsfonts}
\def\iu {\mathrm i}
\def\mb #1{\mbox{\boldmath $#1$}}
\def\rmd {\mathrm d}
\makeatletter
\def\eqnarray{%
   \stepcounter{equation}%
   \def\@currentlabel{\p@equation\theequation}%
   \global\@eqnswtrue
   \m@th
   \global\@eqcnt\z@
   \tabskip\@centering
   \let\\\@eqncr
   $$\everycr{}\halign to\displaywidth\bgroup
       \hskip\@centering$\displaystyle\tabskip\z@skip{##}$\@eqnsel
      &\global\@eqcnt\@ne\hfil$\displaystyle{\hbox{}##\hbox{}}$\hfil
      &\global\@eqcnt\tw@ $\displaystyle{##}$\hfil\tabskip\@centering
      &\global\@eqcnt\thr@@ \hb@xt@\z@\bgroup\hss##\egroup
         \tabskip\z@skip
      \cr
}
\makeatother
\def\rf#1{(\ref{eq:#1})}
\def\lab#1{\label{eq:#1}}
\def\nonu{\nonumber}
\def\br{\begin{eqnarray}}
\def\er{\end{eqnarray}}
\def\be{\begin{equation}}
\def\ee{\end{equation}}

\def\({\left(}
\def\){\right)}

\newcommand{\ct}[1]{\cite{#1}}
\newcommand{\bi}[1]{\bibitem{#1}}

\relax

\def\d{\delta}

\def\h{{1\over 2}}

\def\o{\over}

\def\pa{\partial}

\def\tp0{\Theta_{+}^{(0)}}
\def\tm0{\Theta_{-}^{(0)}}

\def\u2{\mid u\mid^2}

\def\ck{{\cal K}}
\def\cl{{\cal L}}

\newcommand{\IR}{\mathbb R}

\newcommand{\IZ}{\mathbb Z}
\def\one{\hbox{{1}\kern-.25em\hbox{l}}}
\def\0#1{\relax\ifmmode\mathaccent''7017{#1}%
        \else\accent23#1\relax\fi}

\def\PRL#1#2#3{{\sl Phys. Rev. Lett.} {\bf#1} (#2) #3}
\def\NPB#1#2#3{{\sl Nucl. Phys.} {\bf B#1} (#2) #3}

\def\PRD#1#2#3{{\sl Phys. Rev.} {\bf D#1} (#2) #3}

\def\PLB#1#2#3{{\sl Phys. Lett.} {\bf #1B} (#2) #3}

\def\AoP#1#2#3{{\sl Ann. of Phys.} {\bf #1} (#2) #3}

\begin{document}

\begin{titlepage}
\noindent
December, 2000 \hfill{IFT--P.101/2000}\\
hep-th/0012176 \hfill{}

\vskip 3cm

\begin{center}
{\Large\bf Hopf solitons and area preserving\\[.5em]
diffeomorphisms of the sphere}
\vglue 1  true cm
L. A. Ferreira$^1$ and A. V. Razumov$^2$\\

\vspace{1 cm}

$^1${\footnotesize Instituto de F\'\i sica Te\'orica -- IFT/UNESP\\
Rua Pamplona 145\\
01405-900, S\~ao Paulo - SP, Brazil}\\

\vspace{1 cm}
$^2${\footnotesize Institute for High Energy Physics, \\
142284 Protvino, Moscow Region, Russia}\\

\medskip
\end{center}

\normalsize
\vskip 0.2cm

\begin{abstract}
We consider a $(3+1)$-dimensional local field theory defined on the sphere
$S^2$. The model possesses exact soliton solutions with non trivial Hopf
topological charges, and  infinite number of local conserved
currents. We show that the Poisson bracket algebra of the 
corresponding charges is isomorphic to that of the area preserving
diffeomorphisms of the sphere $S^2$. We also show that  the
conserved currents under consideration are the Noether currents associated
to the invariance of the Lagrangian under that infinite group of
diffeomorphisms. We indicate possible generalizations of the model.

\end{abstract}
\end{titlepage}

We consider a Lagrangian field theory model defined on four-dimensional
Minkowski space-time with coordinates $x^\mu$. The fields of the model form
a three-dimensional vector-valued function $\mb n(x)$ satisfying the
relation
\be
\mb n^2(x) = 1. \lab{ns}
\ee
and so, the target space is the two-dimensional sphere $S^2$.
The Lagrangian density is defined as \ct{AFZ99a,kundu}
\be
\cl = - \frac{2}{3} \, \( \frac{H_{\mu \nu} H^{\mu \nu}}{8}
\)^{3/4},
\lab{nicemodel}
\ee
where the tensor $H_{\mu\nu}$ is defined in terms of the three
component unit vector field ${\mb n} \in S^2 $ as
\be 
H_{\mu\nu} \equiv {\mb n} \( \pa_{\mu} {\mb n} \times
\pa_{\nu}{\mb n}\),
\lab{hmn}
\ee
where $\partial_\mu \mb n(x) = \partial \mb n(x) / \partial x^\mu$.

The power $3/4$ of $H_{\mu\nu} H^{\mu \nu}$ in \rf{nicemodel} is such 
that the theory circumvents the usual obstacle of the Derrick's scaling
argument against existence of stable solitons. We are interested in the
boundary condition $\mb n = (0, 0, 1)$ at spatial infinity. 
This condition compactifies effectively the Euclidean space $\IR^3$ to the
three-sphere $S^3$. Accordingly, ${\mb n}$ becomes a map: $ S^3 \to S^2$.
Due to the fact that $\pi_3 ( S^2) = \IZ$, the field configurations fall
into disjoint classes characterized by the value of the Hopf invariant
$Q_H$.

Using the stereographic projection of $S^2$, we introduce instead of the
vector-valued function $\mb n(x)$ satisfying \rf{ns} the scalar complex
field $u(x)$ so that
\be
\mb n = {1\o {1 + |u|^2}} \, \( u + u^*,\, -\iu \, (u - u^*), \,
1 - |u|^2 \). 
\lab{stereo}
\ee
For the tensor $H_{\mu \nu}$ one obtains
\be 
H_{\mu\nu} = {2 \, \iu \o \(1 + |u|^2 \)^2 } \, \( \pa_{\mu}u \,
\pa_{\nu}u^* - \pa_{\nu}u \, \pa_{\mu}u^*\).
\lab{hmnu}
\ee

The Euler-Lagrange equations following from \rf{nicemodel} are
\be
\pa_{\mu}\ck^{\mu} = 0
\lab{eqmot2}
\ee
and its complex conjugate, and where
\be
 \ck^{\mu} = {  1 \o {(1 + |u|^2) \( K \pa u^* \)^{1/4}}} \, K^\mu
\ee
with
\be
K^{\mu} =   
\( \pa_{\nu} u \, \pa^{\nu} u^* \) \, \pa^{\mu} u - 
\(\pa_{\nu} u \, \pa^\nu u \) \, \pa^{\mu} u^*.
\lab{kmu}
\ee

In reference \ct{afz99b} it has been constructed an infinite set of exact
soliton solutions with Hopf topological charge $Q_H = -nm$ and masses
$M_{m,n} \sim \sqrt{|n| \, |m| \(|n| + |m|\)}$, with $m$ and $n$ being
integer numbers. These solitons are similar in some aspects to the solitons
of the Skyrme--Faddeev model \ct{F-S}.
They carry the same topological Hopf charge, and they present knot like
configurations. The basic difference however is that the static solutions
of \rf{nicemodel} have an energy invariant under rescaling of the space
coordinates. The solitons of the Skyrme--Faddeev model, on the other hand,
have a size determined by the balance of the two terms of the Lagrangian
with different scaling properties. In addition, the Skyrme--Faddeev model
is not an integrable theory and its solutions have been found by numerical
methods \ct{numerical}. The Skyrme--Faddeev theory however, contains a
submodel, found in \ct{AFZ99a}, that possesses an infinite number of local
conservation laws. That fact may help the development of exact methods in
the Skyrme--Faddeev theory. Recently \ct{skyrmefsg} it has been found a
similar submodel inside the Skyrme theory \ct{skyrme}. 
 
The integrability properties of  the model described by
the Lagrangian density \rf{nicemodel} have been analyzed in reference 
\ct{AFZ99a}, using the generalized version of the zero curvature condition
\ct{afg}. It has been  shown that such theory possesses an infinite number
of conserved currents given by
\be
J^{\mu}_G = \iu \left[ \ck^{\mu}\, {\partial G \o \partial u} - 
 \ck^{\mu*} \, {\partial G \o \partial u^*} \right],
\lab{infcur2}
\ee
where $G$ is an arbitrary function of $u$ and $u^*$ but
not of their derivatives. We introduced the imaginary unit into relation
above to have a real current for a real function $G$. The conservation of
the currents $J^\mu_G$ is a consequence of the equations of motion
\rf{eqmot2}, and of the fact that the quantity \rf{kmu} automatically
satisfies the relations  
\be
K^{\mu} \pa_{\mu} u = 0 \qquad \qquad {\rm Im}\( K^{\mu} \pa_{\mu} u^*\) =
0.
\lab{nicerel}  
\ee

We calculate in this letter the algebra of the currents $J^\mu_G$ under the
Poisson bracket. The canonical momentum $\pi$ conjugated to $u$ is given
by
\br
\pi = \frac{\partial \cl}{\partial {\dot u}} = \frac{1}{\( 1+|u|^2 \)^3 \(
K \pa u^*\)^{1/4}} \left[ \, \dot u^* \, (\mb
\nabla u \mb \nabla u^*) - \dot u \, (\mb \nabla u^* \mb \nabla u^*) \,
\right]
\, ,
\er
where dot denotes the derivative over $x^0$ and $\mb \nabla$ is the
spatial gradient. The expression for the momentum $\pi^* = \partial \cl/
\partial \dot u^*$ can be obtained by complex conjugation. The
non-vanishing equal time canonical Poisson bracket relations are 
\be
\{ u(\mb x), \, \pi(\mb y)\} = \{ u^*(\mb x), \, \pi^*(\mb y)\} = 
\d (\mb x - \mb y),
\ee
where $\mb x$ and $\mb y$ stand for the three dimensional space
coordinates. 

In terms of the momenta, the time component of the currents take the form
\be
J^0_G = \iu \( 1 + |u|^2 \)^2 \left[ \pi \, \frac{\partial G}{\partial u^*}
- \pi^* \, \frac{\partial G}{\partial u} \right].
\ee
A straightforward calculation shows that
\be
\{ J^0_{G_1}(\mb x), \, J^0_{G_2}(\mb y)\} = J^0_{G_{12}}\(\mb x\) \d (
\mb x - \mb y),
\ee
where
\be
G_{12} = \iu \( 1 + |u|^2 \)^2 \left[ \frac{\partial G_1}{\partial u}
\frac{\partial G_2}{\partial u^*} - \frac{\partial G_1}{\partial u^*}
\frac{\partial G_2}{\partial u} \right].
\ee
For the corresponding charges
\be
\mathcal Q_G = \int \rmd \mb x J^0_G(\mb x)
\ee
we have
\be
\{\mathcal Q_{G_1}, \, \mathcal Q_{G_2}\} = \mathcal Q_{G_{12}}.
\lab{calg}
\ee
Thus, the charges $\mathcal Q_G$ form a closed algebra with respect to the
Poisson bracket.

It can be shown that the algebra under consideration is actually the Lie
algebra of the group of area preserving diffeomorphisms of the
two-dimensional sphere. In order to establish that, let us introduce 
real coordinates $u^1$ and $u^2$ connected with the complex
coordinate $u$ by $u = u^1 + \iu \, u^2$. For the metric tensor components
we have
\be
g_{ab}(u) = \frac{4}{\( 1 + (u^1)^2 + (u^2)^2 \)^2} \,  \delta_{ab}, \qquad
a, b = 1,2.
\lab{metric}
\ee
The area 2-form is defined as
\be
A = \h \, \sqrt{\det g(u)} \, \epsilon_{ab} \, \rmd u^a \wedge
\rmd u^b = \frac{2}{\( 1 + (u^1)^2 + (u^2)^2 \)^2} \, \epsilon_{ab} \rmd
u^a \wedge \rmd u^b. 
\lab{areaform}
\ee
where $g(u) = \|g_{ab}(u)\|$ and $\epsilon_{ab}$ is the totally
skew-symmetric symbol normalized by the condition $\epsilon_{12} = 1$.
The area form $A$ is invariant under a diffeomorphism $\Phi$ of
$S^2$ if
\begin{equation}
\sqrt{\det g(\Phi(u))} \det \left\| \frac{\partial \Phi^a(u)}{\partial
u^b} \right\| = \sqrt{\det g(u)}. \label{7}
\end{equation}
For an infinitesimal diffeomorphism we have
\be
\Phi^a(u) = u^a + \varepsilon X^a(u).
\lab{infdiff}
\ee
The infinitesimal version of relation (\ref{7}) is
\begin{equation}
\partial_a [\sqrt{\det g(u)} \, X^a(u)] = 0. \label{11}
\end{equation}
The general solution of this equation can be written as
\begin{equation}
X^a_F(u) = \frac{1}{\sqrt{\det g(u)}} \, \epsilon^{ab} \,
\frac{\partial F(u)}{\partial u^b},
\label{13}
\end{equation}
where $F$ is an arbitrary function of the coordinates $u^1$ and $u^2$.

Recall that the commutators of the vector fields describing infinitesimal
transformations of some Lie transformation group reproduce the Lie algebra
of the Lie group under consideration. 
In general we have
\begin{equation}
[X_{F_1}, X_{F_2}] = X_{F_{12}}. \label{12}
\lab{vecalg}
\end{equation}
and calculations show that 
\be
F_{12}(u) = - \frac{1}{\sqrt{\det g(u)}} \, \epsilon^{ab} \, \frac{\partial
F_1(u)}{\partial u^a} \frac{\partial F_2(u)}{\partial u^b} \, .
\ee
Using the complex coordinate $u$ we come to
\be
F_{12} = \frac{\iu}{2} \( 1 + |u|^2 \)^2 \left[ \frac{\partial
F_1}{\partial u}
\frac{\partial F_2}{\partial u^*} - \frac{\partial F_1}{\partial u^*}
\frac{\partial F_2}{\partial u} \right]
\ee
and so, the algebras \rf{calg} and \rf{vecalg} are isomorphic 
under the correspondence
$\mathcal Q_{F/2} \leftrightarrow X_F$. Therefore, the algebra of the
charges $\mathcal Q_G$ is indeed the Lie algebra of the group of area
preserving diffeomorphisms of $S^2$. 

Certainly, the established isomorphism of the Lie algebras is not
accidental. As we now show, it is a consequence of the fact that the
Lagrangian is invariant under the area preserving diffeomorphisms of $S^2$,
and consequently \rf{infcur2} are the corresponding Noether currents. 
Notice that using the complex coordinate $u$, we obtain the following 
expression for the area form 
\be
A = \frac{2 \, \iu}{\(1 + |u|^2 \)^2} \, \rmd u \wedge \rmd u^*.
\ee
Using this expression, we see that the 2-form
\br
&& \h \, H_{\mu\nu} \rmd x^{\mu}\wedge \rmd x^{\nu} \nonu \\
&& \hspace{2em} {} = { \iu \o \(1 + |u|^2 \)^2}\,\(\pa_{\mu}u \,
\pa_{\nu}u^* - \pa_{\nu}u
\, \pa_{\mu}u^*\) \, \rmd x^{\mu}\wedge \rmd x^{\nu} =  \frac{2 \, \iu}{\(
1 + |u|^2 \)^2} \, \rmd u \wedge
\rmd u^* 
\lab{pullback}
\er
is the pull-back of the area form on $S^2$. Therefore this 2-form is
invariant with respect to the field transformations induced by
the area preserving diffeomorphism of $S^2$ and any Lagrangian
constructed from $H_{\mu\nu}$ is invariant under those transformations. It
then follows that the currents \rf{infcur2} are in fact the Noether
currents associated with the area preserving diffeomorphisms. 

In general one can consider the $m$-dimensional Minkowski space-time and
take as the target manifold of a model an arbitrary $n$-dimensional
Riemannian manifold $N$ with $n \le m$. The vector fields describing
infinitesimal volume preserving diffeomorphisms of $N$ are given by
\be
X^a_F(u) = \frac{1}{(n-2)!} (\det g(u))^{-1/2} \epsilon^{a a_1 \ldots
a_{n-1}} \partial_{a_1} F_{a_2 \ldots a_{n-1}}(u),
\lab{vpd}
\ee
where $F_{a_1 \ldots a_{n-2}}$ are the components of an $(n-2)$-form $F$ on
$N$. The tensor
\be
H_{\mu_1 \ldots \mu_n} = \frac{1}{n!} \sqrt{\det g(u)} \, \epsilon_{a_1
\ldots a_n} \, \partial_{\mu_1} u^{a_1} \ldots \partial_{\mu_n} u^{a_n}
\ee
is invariant with respect to the field transformation induced by
the volume preserving diffeomorphism of the manifold $N$. Therefore, if we
construct the density of the Lagrangian as an arbitrary function of
\be
H_{\mu_1 \ldots \mu_n} H^{\mu_1 \ldots \mu_n} =
\frac{1}{n!} \det g(u) \det \| \partial_\mu u^a \partial^\mu u^b \|
\ee
we will come to a Lorentz invariant model which is also invariant with
respect to the volume preserving diffeomorphisms of the target manifold
$N$. 

{\it Acknowledgement.} 
The authors are grateful to H. Aratyn, O. Babelon, 
J.F. Gomes, J. S\'anchez Guill\'en, and A. H. Zimerman for discussions. 
One of the authors (A.V.R) wishes to acknowledge the
warm hospitality of the Instituto de F\'\i sica Te\'orica -- IFT/UNESP,
S\~ao Paulo, Brazil, and the financial support from FAPESP during his stay
there in February-July 2000. The research program of A.V.R. was supported
in part by the Russian Foundation for Basic Research under grant \#
98--01--00015,  and L.A.F. was partially suported by CNPq
(Brazil).

\end{document}